# The Roles of Small Telescopes in a Virtual Observatory Environment

S. G. Djorgovski

*Palomar Observatory, Caltech, Pasadena, CA 91125, USA*

**Abstract**: The advent of the Virtual Observatory (VO) concept signals a pradigm shift in the way astronomy will be done in the era of information abundance and ubiquitous networking. Small telescopes will be playing a number of essential roles in this new research environment, probably contributing a major portion of all data taken in astronomy, both as surveying instruments, and as follow-up facilities. In this review we describe the VO concept and its background, and how small telescopes will fit in this emergent way of doing astronomy in the 21$^{st}$ century.

## 1. Introduction: New Astronomy in the Era of Information Abundance

Astronomy, like most sciences, has become immensely data-rich, with data sets measured in many Terabytes, and soon Petabytes. The sky is being surveyed systematically at many wavelengths, with billions of stars, galaxies, quasars, and other objects detected and measured with an unprecedented level of detail. These massive data sets are a new empirical foundation for the astronomy of the 21$^{st}$ century, hopefully leading to a new golden era of discovery.

The steep increase in the volume and complexity of available information is based on the great progress in technology, including digital imaging (the chief data source in astronomy), and, of course, the ways of processing, storing, and accessing information. Most of the scientific measurements and data obtained today are either generated in a digital form, and most instruments contain digital imaging arrays. Such devices are based on the same technology governed by Moore's law, and are thus growing exponentially in their information-generating ability, with the number of bits in astronomy doubling every 1 to 2 years, while the telescope technology develops more slowly (Szalay & Gray 2001).

Large digital sky surveys and archives now becoming the principal sources of data for astronomy. Most of them are generated using small telescopes. In the past, most astronomical studies dealt with individual objects or small samples thereof (tens to hundreds). Now increasingly the field is being dominated by the analysis of large, uniform sky surveys, sampling millions or billions of sources, and providing tens or hundreds of measured attributes for each of them. This produces a selection of optimal targets to follow up with other telescopes, both large and small, and both ground- and space-based. There is a paradigm shift in observational astronomy, with survey-based science becoming an ever more important way of exploring the universe in a systematic way, and leading to the Virtual Observatory astronomy.

The challenges and opportunities posed by individual massive data sets (surveys and archives) are just a start. The sky is now being surveyed over a full range of wavelengths, giving us, at least in principle, a panchromatic and less biased view of the universe. Each individual sky survey has its own intrinsic limitations in terms of the depth, area coverage, angular resolution, etc., but this can be quantified and at least partially overcome by combining different data sets. Moreover, the universe itself imposes selection effects on what can be seen at any given wavelength: for example, obscuration by dust can hide bursts of star formation and active galactic nuclei from the view in the visible or UV light. Some astrophysical phenomena generate the bulk of their luminosity in a particular region of the spectrum, and may be effectively undetectable or appear inconspicuous as seen at other wavelengths. Federation of surveys and data sets spanning a range of wavelengths (or spatial scales, or time scales, or types of observations, e.g., images and spectra, etc.) can help us discover new knowledge which is present in the data, but not apparent in any of the data sets viewed individually. Joining of massive data sets into even larger and more complex ones thus becomes a scientific imperative.

While the quantity and complexity (and quality) of data continue to grow exponentially, our knowledge and understanding of the world has not kept pace: *we are not making the full use of the information richness available to us.* Transforming vast masses of bits into a refined knowledge and understanding of the universe is a highly complex task. The problems go well beyond technical, and touch the very core of scientific methodology and practice. The great quantitative change in the amount and complexity of available scientific information should lead to a qualitative change in the way we do science.

Old research methodologies, geared to deal with data sets many orders of magnitude smaller and simpler than what we have now, are becoming inadequate. We have to develop new scientific methodologies which would maximize the utility, effectiveness, and scientific returns of the ever increasing, massive data sets, and enable us to exploit the great opportunities in front of us.

## 2. The Virtual Observatory Concept

In order to cope with these challenges, the astronomical community started an initiative, the National (and ultimately Global) Virtual Observatory (VO). The VO would federate numerous large digital sky archives, provide the information infrastructure and standards for ingestion of new data and surveys, and develop the computational and analysis tools with which to explore these vast data volumes.

Recognizing the urgent need, the National Academy of Science Astronomy and Astrophysics Survey Committee (McKee, Taylor, *et al* 2001) recommends the establishment of a National Virtual Observatory (NVO) as a first priority in their "small" (i.e., cost < $ 100 M) project category. The early vision of the NVO was presented in the NVO White Paper (2001). In order to provide further community input, to refine the scientific drivers and technological requirements, and draft a roadmap for the NVO, NASA and the NSF formed the NVO Science Definition Team, whose report (NVO SDT 2002) summarizes the relevant issues.

This initiative is rapidly growing and gathering support worldwide. Several major international conferences and topical sessions at general conferences have been dedicated to it; see for example, the volumes edited by Brunner *et al.* (2001) and Banday *et al.* (2001), several dedicated sessions at SPIE and ADASS meetings, etc. Several major projects are already under way (see the Web Resources at the end of this Chapter for the relevant links). In what follows, we will assume a trans-national VO concept, although of course any of the regional VO's would play essentially the same roles and collaborate on the global level.

It is important to clarify what the VO is and is not. It is not yet another data archive, or a data center, or a digital library of already produced results, although it will contain links to all of such resources. It rests on the solid and highly successful foundations of the existing data centers, archives, and observatories, but goes well beyond them. It is *a novel type of a scientific research organization* for the Internet era. It has to be geographically distributed, since the data, the expertise, and the computational resources are spread world-wide and are evolving at a rapid pace. It transcends the traditional divisions and agency domains based on the wavelength, observatory- or mission-specific, ground- vs. space-based, etc. The VO will be much more than a depository of massive data sets; it will provide both data-oriented services (archiving and metadata standards, interfaces, protocols, etc.) and data analysis tools and capabilities (survey federation, data mining, statistical analysis, visualization, observations-theory comparisons, etc.). It will have an unprecedented, broad range of users and constituents. *It will be a complete, web-based, distributed research environment for astronomy with massive data sets.* Ultimately, it will become the very fabric of the astronomical practice and become a part of the invisible infrastructure.

The VO will provide new opportunities for scientific discovery that were unimaginable just a few years ago. Entirely new and unexpected scientific results of major significance will emerge from the combined use of the resulting datasets, science that would not be possible from such sets used singly. *The VO will serve as an engine of discovery for astronomy.* At the same time, it will provide great returns to the partnering disciplines, e.g., the applied computer science and information technology (CS/IT), statistics, etc.

The VO concept also represents *a powerful engine for the democratization of science.* Making first-rate resources (the data and the tools for their exploration) available to the entire astronomical community via the Web would engage a much broader pool of talent, including scientists and students from small educational institutions in the U.S. and from countries without access to modern or large telescopes or space observatories, but with strong intellectual traditions. Human intelligence and creativity are now distributed much more widely than the technological resources of the cutting-edge astronomy (or indeed any science). Who knows what great ideas would emerge from this pool? This opening of a scientific opportunity to everyone with a Web access may have a very significant impact in the long run.

Finally, the VO offers unprecedented opportunities for education and public outreach and involvement at all levels: from pre-school to college, including the amateur astronomy community, the media, etc.

## 3. From the Survey–Based Astronomy to the VO–Based Astronomy

These developments are already changing the way in which the observational astronomy is done. There is a great increase in the efficiency of use of our most powerful facilities, e.g., large telescopes from the ground or orbiting space observatories. In the past, samples of interesting objects to study have been selected using such expensive facilities themselves; this is demonstrably a poor use of such valuable resources. Now, the most promising targets are selected from large sky surveys, and then followed up with the more expensive resources like the 8-to-10-meter class telescopes or the HST (a small, but highly effective telescope in space). Examples include discoveries of copious numbers of brown dwarfs by the SDSS and 2MASS, and discoveries of large numbers of distant quasars by DPOSS and SDSS, including the most distant objects now known; their follow-up studies are already providing us with valuable new insights into the early universe and structure formation.

The abundance of uniform, high-quality data from large, digital sky surveys is already transforming some fields of astronomy. The nascent field of precision cosmology uses large samples of galaxies and quasars from surveys like SDSS or DPOSS to probe the large-scale structure of the universe and its evolution with unprecedented accuracy. Comparable changes are expected in Galactic astronomy, providing a quantitative description of the structure and dynamics of our own Galaxy at a level of detail never seen before, including a slew of new insights into the stellar astrophysics, the distribution of the dark matter in the universe, etc. These efforts have demonstrated that the old data analysis methodology in astronomy, geared to deal with data sets measured in Megabytes to Gigabytes and samples of objects measured in hundreds and thousands, is not adequate to deal with multi-Terabyte data sets and catalogs of many millions or billions of sources.

The mode of the survey-based astronomy is illustrated schematically in Figure 1. A dedicated (typically small) telescope is generating the data, which are processed and stored in a survey archive. Scientific exploration of this archive through various data mining techniques generates scientific results based on the survey data alone; for example, one can study the large-scale structure in the universe from panoramic imaging alone. In addition, interesting sources are found and followed up spectroscopically or in other ways using other observational facilities (large telescopes, space observatories, etc.); examples may include quasar or brown dwarf candidates, various variable sources, complete samples of galaxies for massive redshift surveys, etc.

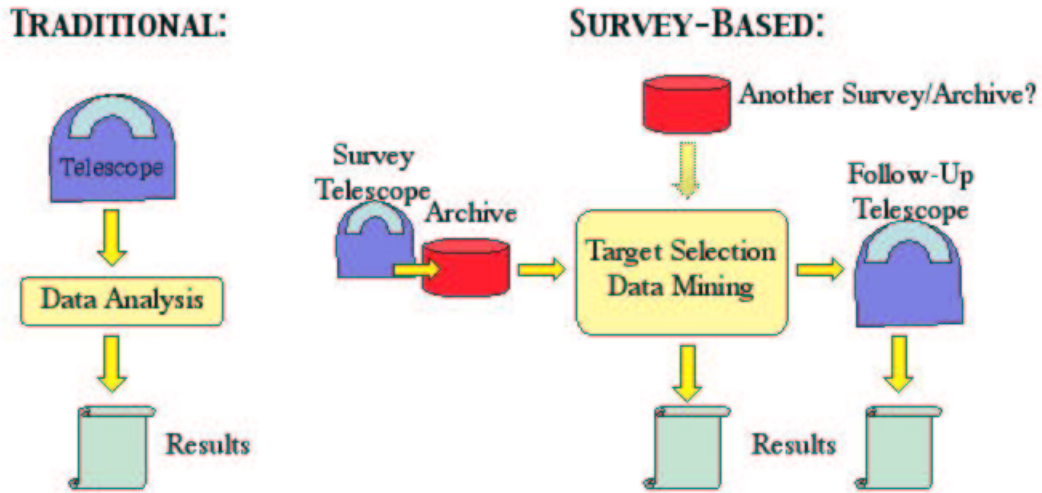

*Figure 1*. A schematic outline of the traditional and the survey-based astronomy

Survey-based astronomy is coming of age and producing excellent science. The natural question is, what next? Discoveries made using individual survey, while interesting and valuable, are constrained by the information content of these surveys. There is a lot of added value to be had from data fusion, from federating of the multiple sky surveys to gain a more comprehensive view of the universe. What comes next, beyond the survey science, is the Virtual Observatory science.

Every astronomical observation and every survey covers a portion of the observable parameter space, whose axes include the area coverage, wavelength coverage, limiting flux, etc., and with a limited resolution in angular scales, wavelength, temporal baseline, etc. Each one represents a partial projection of the observable universe, limited by the observational or survey parameters (e.g., pixel sampling, beam size, filters, etc.). Every astronomical data set samples only a small portion of this grand observable parameter space, usually covering only some of the axes and only with a limited dynamical range along each axis. Every survey is also subject to its own selection and measurement limits.

Surveys thus represent hypervolumes in the observable parameter space. Individual sources represent data points (or vectors) in this multidimensional parameter space. So far we have sampled well only a relatively limited set of sub-volumes of this observable parameter space, with a much better coverage along some of the axes than others. Some limits are simply technological or practical, but some are physical, e.g., the quantum noise limits, the opacity of the Earth's atmosphere, or the Galactic interstellar medium.

Federating multiple surveys that sample different portions of the observable parameter space can provide a much more complex and complete view of the physical universe. The simplest and most traditional, yet very powerful, manifestation of this process is the cross-identification of observations at different wavelengths.

Historical examples abound. The discovery of quasars resulted when some of the first well-localized radio sources were identified in the visible light with what appeared as otherwise non-descript bluish stars – yet they represented a new, spectacular natural phenomenon, and the science of astronomy was changed fundamentally. A similar revelation happened when the sky was mapped for the first time in the far-infrared by the IRAS satellite: it was discovered that the most luminous objects in the nearby universe, identified optically with some irregular galaxies previously considered as mere curiosities, have their powerful energy sources hidden behind the veils of opaque dust. It is now believed that at least a half of all star formation and a large fraction of all AGN in the universe are escaping detection in the standard visible-light techniques due to such obscuration. Cosmic gamma-ray bursts (GRBs) are another spectacular example, where the breakthrough in the understanding of a new astrophysical phenomenon came from insights gained at other wavelengths, including x-ray, optical, and radio: a 30-year cosmic mystery was resolved by combining the data from a range of wavelengths. Essentially all of the x-ray astronomy illuminated the universe in a new light: clusters of galaxies are primarily x-ray emitting objects, and we have gained great insights into a range of variable stars and AGN from the x-ray point of view, with information not available in any other wavelength regime.

These remarkable discoveries represent simple examples of what is possible when data from a range of wavelengths are combined in a systematic way. *Fundamental new phenomena are found, and new insights are gained in the types of objects and processes already known to exist*. We can surely expect even more spectacular discoveries and insights as we start to combine and explore in detail the new generation of massive digital sky surveys, leading us to a better, more complete and deeper understanding of the physical universe.

This synergy of data from many different sources (many of which are generated with small telescopes), illustrated schematically in Figure 2, is a more complex, but also scientifically much more rewarding process, which allows us to fully exploit the information content of the data sets, which we could not do with any of them individually. Follow-up studies are just as important as in the case of survey-based astronomy, and small telescopes can play a significant role here, especially for the photometric follow-up of variable sources found in synoptic sky surveys. The VO thus plays a role of a unifying and enabling environment for astronomy with a range of facilities and telescopes, from small to large.

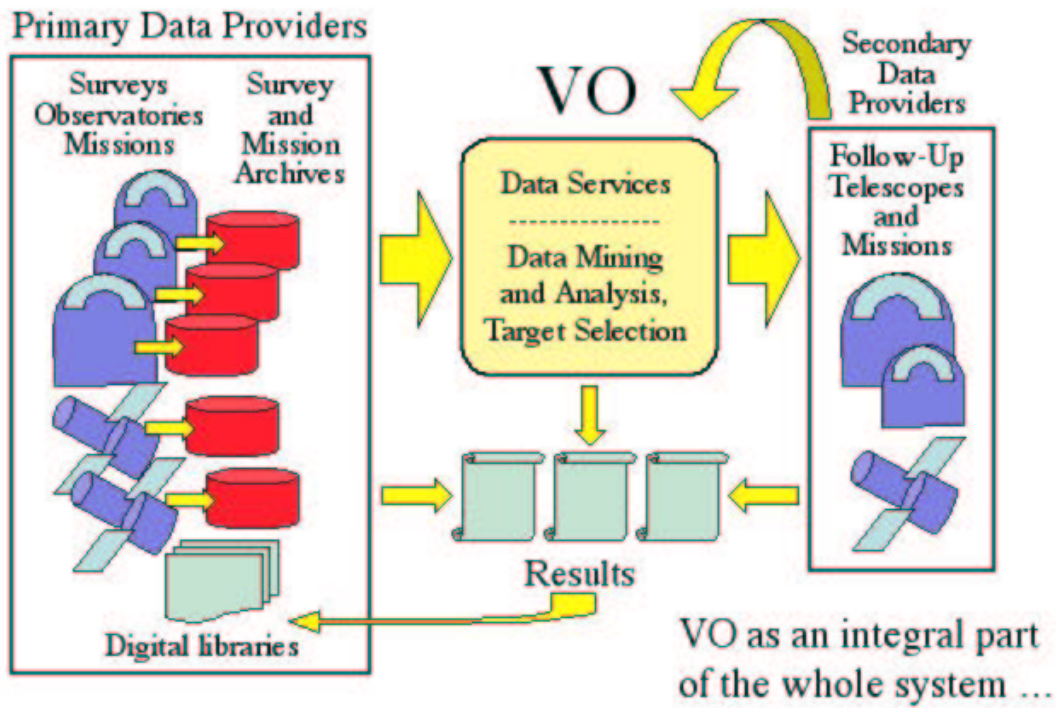

*Figure 2.* A Schematic Outline of VO-Based Astronomy

A parameter space representation of the available data shows us not only what is known about the universe, but also what are the unexplored areas where genuine new discoveries may be made, and also what is knowable given the technological or physical limits of our measurements (see also Harwit 1975). *Now for the first time we have the adequate data and technology for such a global, empirical approach to the exploration of the universe.* This has never been possible or done before.

A typical VO data set may be a catalog of $\sim 10^8$–$10^9$ sources with $\sim 10^2$ measured attributes each, i.e., a set of $\sim 10^9$ data vectors in a $\sim$ 100-dimensional parameter space. Objects of the same physical type would form clusters or sequences in this parameter space. The most populous clusters will be those containing common types of objects, e.g., normal stars or galaxies. But the unusual and rare types (including possible new classes) would form sparse clusters or even be just individual outliers, statistically distinct from the more common ones. Rare objects may be indistinguishable from the more common varieties in some observable parameters, but be separable in other observable axes. This approach has been proven to work and is now used on an industrial scale to find high-redshift quasars and brown dwarfs in SDSS, 2MASS, or DPOSS.

A typical VO project may be a search for the previously unknown types of objects or rare phenomena associated with known types, manifesting themselves through "anomalous" properties, i.e., as "outliers" in some large parameter space. If some type of an interesting object is, for example, one in a million or a billion down to some flux limit, then we need a sample of sources numbering in many millions or billions in order to discover a reasonable sample (or even just individual instances) of such rare species.

Another approach, where we may expect more novelty and surprises, is a systematic exploration of the poorly known portions of the observable parameter space. One such domain is the variability on the sky, especially at faint flux levels, at every wavelength. While a number of variable types of objects are already known, including many types of variable stars, quasars, novae and supernovae, GRBs, etc., we know very little about the time-variable universe in a systematic way (see, e.g., Paczynski 2000). There are already some puzzling phenomena found, e.g., the fast, faint optical transients, which may or may not be associated with distant supernovae, and the mega-flares on otherwise apparently normal stars, which brighten by a factor of a few hundred for a period of hours or days, for as yet unknown reasons.

A new generation of synoptic digital sky surveys, which will image repeatedly large portions of the sky down to very faint limits, will generate the necessary data to start exploring the time domain. A number of NASA-sponsored groups are patrolling the sky with small telescopes, searching for moving objects, and in particular Earth-crossing ("killer" or extinction-causing) asteroids. The same data sets can be also mined for variable objects.

## 4. Conclusion: A Synergy of Facilities

As the practice of astronomy undergoes a profound change within the VO paradigm, small telescopes will be among the key constituents of the overall system, fulfilling the essential roles for which larger telescopes are simply not suitable or appropriate. (This is not to say that large telescopes do not play a significant role – they certainly do, especially, but not only for spectroscopic observations.)

First, small telescopes will continue to be the facilities where large sky surveys are generated, especially synoptic surveys, which will open the very exciting new frontier of exploration of the time domain. Second, they can be essential facilities for a rapid, coordinated follow-up of interesting objects and phenomena, again especially in the time domain. These surveys are expected to generate large numbers of viable targets, for which time-critical observations would be essential. Rapid follow-up observations of highly variable or transient sources discovered at any wavelength (from γ-rays to radio) would be essential in understanding and classification of such objects (e.g., a supernova, a GRB afterglow, an OVV AGN, a CV star, … or something really new). While the VO would provide the analysis environment, the data would have to come from somewhere, and in most cases that may be small telescopes.

Two key issues along this path are telescope dedication and automation. It makes little scientific and operational sense to have universally equipped small telescopes; their scientific and cost effectiveness can be vastly higher if they are dedicated to specific kinds of data gathering, with specific instruments (e.g., a multi-bandpass CCD or IR camera). This is self-evident in the case of survey telescopes, but it is also valid for the follow-up facilities. Likewise, automated operation of such telescopes seems to be a practical necessity. The only exceptions to these requirements may be the telescopes with a strong educational, hands-on component. In many cases it may make sense to form *networks of small telescopes* dedicated for a particular purpose; an example may be the follow-up of GRB observations, e.g., as in the REACT collaboration (http://react.srl.caltech.edu). In any case, the large data volumes and the need for rapid reactions imply that such telescopes must be connected to a proper information infrastructure.

Thus, the future of small telescopes in the VO era seems bright and assured, as the key data providers and as rapid reaction probes of the new and exciting phenomena in the new era of information-rich astronomy.

**Acknowledgements:** I wish to thank numerous colleagues and collaborators who contributed to the development of ideas and concepts described here, and in particular to R. Brunner, A. Mahabal, A. Szalay, T. Prince, R. de Carvalho, R. Gal, S. Odewahn, R. Williams, and many others. This work was supported in part by grants from the NSF and NASA, and benefited from the creative environment of the Aspen Center for Physics.

**Selected Web Resources:**

The NVO SDT Website: http://www.nvosdt.org

The US NVO ITR Project: http://us-vo.org

The VO Forum: http://voforum.org/

The European AVO Project: http://www.eso.org/projects/avo/

The UK Astrogrid Project: http://www.astrogrid.ac.uk

The author's VO Webpage: http://www.astro.caltech.edu/~george/vo/

**Selected Current and Forthcoming Digital Sky Surveys:**

Digital Palomar Observatory Sky Survey: http://dposs.caltech.edu

Two-Micron All-Sky Survey: http:// www.ipac.caltech.edu/2mass/

Sloan Digital Sky Survey: http://www.sdss.org

The ESO VST Project: http://www.na.astro.it/vst/

The UK VISTA Project: http://www.vista.ac.uk/

A good set of links on synoptic sky surveys with small telescopes is maintained by B. Paczynski at: http://www.astro.princeton.edu/